\documentclass[conference]{IEEEtran}
\usepackage[T1]{fontenc}
\usepackage{cite,amsmath,amssymb,graphicx,booktabs}
\usepackage{algorithm,algpseudocode,tikz}
\usepackage{optidef}
\usetikzlibrary{positioning,arrows.meta}
\usepackage[hidelinks]{hyperref}
\DeclareMathSizes{8}{8}{7}{5}
\begin{document}
\title{EchoWM: Learning Application Feedback for\\Rate-Preserving Carrier Sleep Control}
\author{
    \IEEEauthorblockN{
        Loc~X.~Nguyen\textsuperscript{1}, Ji Su Yoon\textsuperscript{1}, Phuong-Nam Tran\textsuperscript{1}, Huy Q. Le\textsuperscript{2},
        Eui-Nam Huh\textsuperscript{1},\\
        Zhu Han\textsuperscript{3},
        and Choong Seon Hong\textsuperscript{1}\textsuperscript{*}
    }
    \IEEEauthorblockA{
        \textsuperscript{1} School of Computing, Kyung Hee University, Yongin-si 17104, Republic of Korea\\
        \textsuperscript{2}G-LAMP Nexus Institute, Kyung Hee University, Yongin-si, 17104, Republic of Korea\\
        \textsuperscript{3}Electrical and Computer Engineering Department, University of Houston, Houston, TX 77004
    }
    \vspace{-11mm}
}
\maketitle
\begin{abstract}
Carrier sleep has been considered a promising approach to reduce the energy consumption of base stations (BSs) when part of their capacity is unused. However, putting the secondary carrier to sleep reduces the available capacity, and adaptive video clients respond to this reduction by requesting lower bitrates. As a result, the BS observes less traffic even though the service quality declines. To overcome this challenge, we propose EchoWM, a structured world model that captures this action-dependent application response for rate-preserving carrier sleep control. Specifically, EchoWM combines the known capacity, buffer, and energy mechanics of the cell with a neural network that learns the rate adaptation of clients. When client reports are delayed or lost, EchoWM reconstructs the present client state from the latest report and the intervening carrier commands, and then a receding-horizon planner evaluates candidate sleep schedules. Simulation results show that with limited real interactions, EchoWM achieves the highest mean energy saving among the learning-based controllers under the rate and stall limits, i.e., 23.83\% compared with 15.83\% for a deep Q-network (DQN). With more interactions, DQN catches up with and slightly surpasses EchoWM, while a tuned buffer rule saves more energy under the nominal report delay. However, without retuning, the buffer rule violates the rate-loss limit when the report delay becomes longer, whereas EchoWM preserves its service. These results indicate that structured application modeling is beneficial when real interactions are limited and client feedback is delayed.
\end{abstract}
\vspace{-0.1in}
\vspace{-0.1in}
\section{Introduction}\label{sec:intro}
Base stations (BSs) in cellular networks are designed to handle high traffic demand, while the capacity required by their users varies throughout the day~\cite{lowenergy1}. As a result, keeping all carriers active during low-demand periods consumes energy, even though much of their capacity remains unused~\cite{lowenergy2}. To reduce this energy consumption, one promising approach is to keep a primary carrier active for the basic service, while allowing a secondary carrier to sleep when its additional capacity is not needed~\cite{lowenergy0}. However, putting the secondary carrier to sleep reduces the available capacity of the BS, which may degrade the service provided to users~\cite{lowenergy0,lowenergy1,lowenergy2}. Therefore, the \emph{main challenge of carrier sleep control is to determine when the secondary carrier can sleep without degrading the user service}.

In the literature, several approaches have been proposed to improve the energy efficiency of sleep control. The most straightforward strategy is to activate or deactivate radio resources according to the measured traffic load, whereas traffic forecasting allows the controller to anticipate the future demand. For example, the authors in~\cite{deepbsc} proposed DeepBSC, which combines spatiotemporal traffic forecasting with reinforcement learning (RL) to determine the sleep decisions of BSs. In~\cite{kairos}, Kairos coordinates radio scheduling and advanced sleep modes, where a learning-based controller reduces the energy consumption of radio units while satisfying the service requirements. More recently, network world models have been adopted to predict how the network states and performance evolve under different control actions~\cite{bouton,mobiworld,twm}. With these predictions, the controller can evaluate the consequences of energy-saving decisions before applying them to the network.

However, the traffic observed at the BS does not always reflect the user demand, since the traffic itself changes in response to the network control decisions. For instance, adaptive video clients adjust the bitrate of the requested video segments according to the available throughput and playback buffer~\cite{SODA,EnhancingQoE}. When the secondary carrier is put to sleep, the reduced capacity may cause the clients to request lower-bitrate segments, and consequently, the BS observes less traffic even though the users continue watching their videos. If the controller interprets this decrease as a reduction in user demand, it may keep the carrier asleep without recognizing the degradation in video quality. In this case, part of the energy saving comes from delivering lower-quality videos, instead of deactivating the unnecessary capacity.

Moreover, this feedback is difficult to detect from the current traffic measurements due to the playback buffer. To be specific, when the carrier is put to sleep, a client with sufficient buffered content can continue playing the video despite the reduced capacity. However, if the reduction persists, the changes in buffer occupancy and estimated throughput may cause the client to request lower-bitrate segments to avoid playback interruptions~\cite{chen2026}. Therefore, once the clients have adapted, the BS may observe both lower traffic and uninterrupted playback, even though the requested video quality has declined. This observation motivates us to predict how the clients will respond to a proposed capacity reduction, rather than relying only on the traffic observed under the current network configuration.

World models offer a natural way to address this problem, since they can simulate how a system dynamically changes under different control actions and evaluate candidate decisions before executing~\cite{ICLR2024_ece182f9,nguyen2026self}. Recent wireless studies have explored this capability for energy-efficient sleep control~\cite{bouton,mobiworld}. However, they have not explicitly modeled how the applications of users react to the control decisions. Therefore, in this paper, we propose \emph{EchoWM}, a structured world model that captures the application response for carrier sleep control. Specifically, EchoWM combines explicit capacity allocation, playback-buffer evolution, and carrier energy accounting with a neural network that learns the rate adaptation of clients. Since each candidate sleep schedule changes the predicted service and client requests, a receding-horizon controller can use the resulting trajectories to reduce the energy consumption while limiting rate degradation and playback interruptions. Furthermore, when client reports are delayed or missing, EchoWM uses the same learned transition to reconstruct the present application state from the last report and the carrier commands issued afterward. To the best of our knowledge, this is the first study that incorporates the rate adaptation of video clients into a world model for carrier sleep control. The main contributions of this paper are summarized as follows:
\begin{itemize}
\item First, we consider a two-carrier cell that serves adaptive video clients with heterogeneous adaptation profiles, where the client-state reports are delayed or missing. Then, we formulate a carrier sleep control problem that maximizes the energy saving of the cell while preserving the requested rate and playback availability of the clients. 
\item Secondly, we propose EchoWM, which couples the known service and buffer mechanics with a learned client-response function. To handle stale reports, EchoWM replays the observed carrier commands from the latest report to reconstruct the present client state, and then predicts the outcomes of candidate sleep schedules through the application feedback loop.
\item Finally, we conduct a series of simulations to evaluate the energy saving of our proposed EchoWM under the declared rate and stall limits at three real-interaction budgets, and compare it against always-on operation, tuned utilization and buffer rules, deep Q-network (DQN), proximal policy optimization (PPO), and recurrent PPO. 
\end{itemize}

\section{Related Works}\label{sec:related}
\subsection{Learning-Based Energy Management}\label{sec:rel-energy}
In this subsection, we provide an overview of the literature on learning-based energy management, which aims to improve the energy efficiency of networks under varying traffic and service requirements. The authors in~\cite{deepbsc} considered a cellular network with multiple BSs and overlapping coverage areas, and proposed DeepBSC to determine the sleep decisions of BSs. Specifically, DeepBSC uses a geographical and semantic spatiotemporal network to forecast the traffic demand, followed by actor-critic RL to balance the energy consumption, service delay, and switching costs. Unlike~\cite{deepbsc}, Kairos~\cite{kairos} focused on the short idle periods of radio units in open radio access networks (O-RAN). It combines advanced sleep modes with radio scheduling to consolidate transmissions into fewer time slots, while a learning-based xApp adjusts the sleep policy according to the traffic conditions and the quality-of-service (QoS) requirements of network slices. For energy provisioning in AI-enabled radio access networks (AI-RAN), Gain \textit{et al.}~\cite{gain} proposed a ReVIN-Transformer to predict the throughput, power consumption, and block error rate. These predictions are then combined with a lightweight surrogate model and constrained differentiable optimization to adjust the radio parameters for improved energy efficiency. At the carrier level, Elsayed \textit{et al.}~\cite{lowenergy0} adopted RL for energy-efficient component carrier activation and deactivation in 5G networks, while Maggi \textit{et al.}~\cite{lowenergy2} applied Bayesian learning to carrier shutdown to save energy under performance constraints. Furthermore, the authors in~\cite{lowenergy1} modeled the transfer of users during dynamic carrier shutdown in green 5G networks.

\subsection{World Models for Network Control}\label{sec:rel-wm}
In this subsection, we summarize the studies on world models and their applications to network control. World models learn the representations of an environment and its dynamics, which allows agents to predict future states under different actions and supports planning and policy learning through imagined trajectories~\cite{worldmodels,ICLR2024_cf73d57b}. For example, DreamerV.3~\cite{dreamer} uses recurrent latent dynamics and actor-critic learning to improve decision-making across diverse control tasks. For mobile-network energy control, the authors in~\cite{bouton} combined an action-conditioned network model with uncertainty-aware receding-horizon planning to evaluate cell sleep decisions under throughput constraints. Meanwhile, MobiWorld~\cite{mobiworld} adopts conditional diffusion to generate traffic, user distributions, and signal-strength measurements for multi-cell sleep and user offloading. The Telecom World Model framework in~\cite{twm} separates controllable network configurations from external dynamics and proposes field and control models, which are illustrated through network slicing. In millimeter-wave (mmWave) vehicle-to-everything (V2X) networks, Wang \textit{et al.}~\cite{dualmind} combined recurrent state-space dynamics with neural logic to predict consistent future network states and optimize link scheduling for completeness-weighted age of information.

\subsection{Adaptive Bitrate Streaming}\label{sec:rel-abr}
In this subsection, we review the literature on adaptive bitrate (ABR) streaming, in which the clients change their requested video quality in response to the recent throughput and playback buffer. FESTIVE~\cite{festive} uses stateful rate updates to improve the fairness, efficiency, and stability of HTTP-based adaptive streaming, while BBA~\cite{bba} selects the video rate based on the buffer occupancy and was evaluated in a large video streaming service. In~\cite{abrmpc}, the authors proposed a control-theoretic, model-predictive ABR approach that forecasts the future service to select the video segments. Later studies further improved the throughput prediction and adopted learning-based players to respond to changing service conditions. Specifically, the authors in~\cite{bpa2026} enhanced adaptive video streaming through bandwidth prediction with deep RL, and the work in~\cite{eieabr2025} exploited environment information to improve neural ABR streaming for intercity railways. Similarly, EAStream~\cite{eastream2026} is an environment-aware ABR algorithm for reliable video streaming services, and the authors in~\cite{aar2025} combined enhanced bandwidth measurement with robust rate adaptation for low-latency live streaming. However, these studies control the video player at the client side, whereas EchoWM controls the BS carrier and predicts how clients' responses to each carrier command change future traffic.

\subsection{Positions in the Literature and Rationale for Proposal}\label{sec:rel-rationale}
The studies in Sections~\ref{sec:rel-energy} and~\ref{sec:rel-wm} have shown the benefits of learning-based control and world models for network energy management, where the controller anticipates the traffic demand or evaluates the consequences of sleep decisions before applying them. However, these studies do not explicitly model the application-level feedback of video clients, i.e., how a capacity reduction changes the subsequent traffic requested by the clients. On the other hand, the ABR studies in Section~\ref{sec:rel-abr} adapt the requested bitrate at the client side, but they regard the network service as a given condition, whereas this service is determined by the carrier decisions of the BS in our problem. In response, our proposed EchoWM addresses the feedback loop in which carrier sleep changes the delivered service and thereby affects the playback buffers and subsequent video requests. To capture this loop, EchoWM combines explicit service and buffer equations with a learned client-response model, so that the controller can predict the application response to each candidate sleep schedule. Additionally, we consider delayed and missing client reports, and EchoWM reconstructs the present client state by replaying the carrier commands issued since the latest report.

\section{System Model and Problem Formulation}\label{sec:system}
\subsection{Network Model}\label{sec:network}
We consider a single-cell network, which consists of one BS and a set $\mathcal{U}$ of $U$ long-lived streaming clients. The BS operates two carriers: a primary carrier (PCell), which always remains active, and a secondary carrier (SCell), which can be put to sleep. At each interval $t$ with a duration of $\Delta=1$~s, the controller chooses the carrier command $a_t\in\{0,1\}$, where $a_t=0$ puts the SCell to sleep and $a_t=1$ activates it. In this paper, we assume that the previous carrier command is known to the controller. Since reactivating the SCell consumes a fraction $\omega$ of the interval, the available fraction of the SCell can be expressed as follows:
\begin{equation}
v_t=a_t[1-\omega(1-a_{t-1})].
\label{eq:available}
\end{equation}
Accordingly, the aggregate service capacity of the cell is calculated as follows:
\begin{equation}
C_t=\zeta(C_{\mathrm P}+v_tC_{\mathrm S}),
\label{eq:capacity}
\end{equation}
where $C_{\mathrm P}$ and $C_{\mathrm S}$ denote the nominal capacities of the PCell and SCell, respectively, and the factor $\zeta$ represents the effective radio condition. The value of $\zeta$ is fixed within an episode, measured by the controller, and independently varied between episodes. With the abstraction, we can keep the radio propagation separate from the application-feedback problem, which is the main focus of our paper.

Each client $u\in\mathcal{U}$ has a buffer $b_{u,t}$, a requested video rate $R_{u,t}$, a recent-service estimate $m_{u,t}$, and a static adaptation profile $g_u$. The profile is registered as client metadata, and thus, it is visible to every controller. In contrast, the three dynamic values of all clients are carried in a cell-level report, which reaches the controller with a delay of four intervals and is independently dropped with a probability of $\alpha$. To bound the information age, a report is forced through when its age reaches $M$ intervals. Besides the delayed reports, the controller knows the report age (i.e., the number of intervals elapsed since the source interval of the latest delivered report), all carrier commands issued since that report, the current carrier state, and the radio-capacity factor. EchoWM uses its learned transition to propagate the last received client state through the known command history before planning the next action, as described in Section~\ref{sec:proposed}.

\begin{figure}[t]
\includegraphics[width=0.5\textwidth]{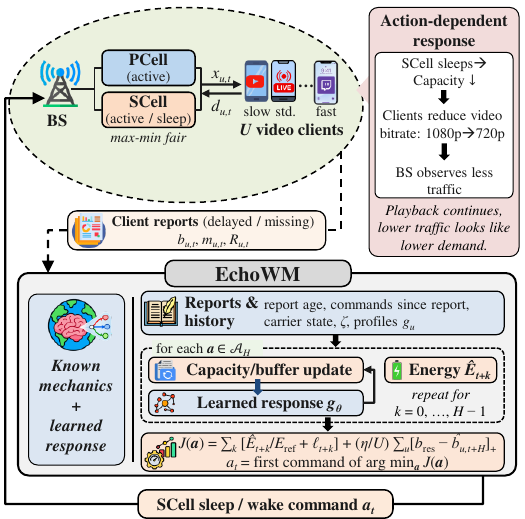}
\caption{Overview of EchoWM, which models application feedback to predict client responses and optimize carrier sleep decisions.}
\label{fig:framework}
\vspace{-0.15in}
\end{figure}

\vspace{-0.1in}
\subsection{Video Streaming and Buffer Model}\label{sec:streaming}
We adopt a fluid streaming abstraction with continuous rate adaptation~\cite{8765424}, in which each client requests enough data to sustain its playback and replenish its buffer toward the target level $B$. Therefore, the offered rate of client $u$ is given by:
\begin{equation}
d_{u,t}=R_{u,t}\operatorname{clip}\!\left(1+\frac{B-b_{u,t}}{\tau_f},\beta\right),
\label{eq:demand}
\end{equation}
where $\tau_f$ controls the refill speed. The scheduler at the BS applies max-min fair sharing among the clients, i.e., $x_{u,t}=\min(d_{u,t},\nu_t)$, where $x_{u,t}$ denotes the service rate delivered to client $u$, and the water level $\nu_t$ satisfies the following condition~\cite{752159}:
\begin{equation}
\sum_u x_{u,t}=\min\left(C_t,\sum_u d_{u,t}\right).
\label{eq:sharing}
\end{equation}

In this way, the unused shares of some clients are redistributed among the other clients. Since the delivered data add $\Delta x_{u,t}/R_{u,t}$ seconds of playable content to the buffer of client $u$, the buffer and the unavailable-playback duration are updated as follows:
\begin{align}
b_{u,t+1}&=\operatorname{clip}\!\left(b_{u,t}+\frac{\Delta x_{u,t}}{R_{u,t}}-\Delta,0,B\right),\label{eq:buffer}\\
s_{u,t}&=\left[\Delta-b_{u,t}-\frac{\Delta x_{u,t}}{R_{u,t}}\right]_+,
\label{eq:stall}
\end{align}
where $[\cdot]_+=\max(\cdot,0)$. Meanwhile, the recent-service estimate of the client is updated by an exponential moving average as follows:
\begin{equation}
m_{u,t+1}= \gamma m_{u,t}+(1-\gamma)x_{u,t}.
\label{eq:memory}
\end{equation}

In general, Eq.~\eqref{eq:available}-\eqref{eq:memory} define the known part of the world model. After receiving the service, each client chooses its next requested rate through the following adaptation function:
\begin{equation}
R_{u,t+1}=f_{\rm client}(b_{u,t+1},m_{u,t+1},R_{u,t};g_u).
\label{eq:client}
\end{equation}
Here, Eq.~\eqref{eq:capacity}-\eqref{eq:client} form the feedback loop of interest, in which a sleep decision changes the delivered service, the service changes the application state, and the application state in turn changes the subsequent offered traffic. Since $f_{\rm client}(\cdot)$ is unknown to the controller, EchoWM learns this response from the observed transitions and relies only on the learned approximation during planning.
One noteworthy point is that the fluid abstraction treats the video data as a continuous flow, instead of a queue of individually encoded segments. Therefore, we measure the \emph{requested video rate} and the playback availability separately, where the requested rate is used as a quality-related operating metric, rather than a reconstructed segment-level playback bitrate or a perceptual quality score.

\subsection{Energy Consumption Model}\label{sec:energy}
The PCell consumes a constant power $P_{\mathrm P}$, while the SCell consumes $P_{\mathrm A}$ when it is commanded active (including its wake interval) and $P_{\mathrm L}$ when it sleeps. In addition, each change of the carrier command incurs a switching energy $E_{\rm sw}$. Following prior sleep-control formulations that account for both
state-dependent operating energy and switching overhead
\cite{deepbsc}, we use the following simplified two-state
energy accounting model:
\begin{equation}
E_t=\Delta[P_{\mathrm P}+a_tP_{\mathrm A}+(1-a_t)P_{\mathrm L}]
+E_{\rm sw}\mathbf1\{a_t\ne a_{t-1}\},
\label{eq:energy}
\end{equation}
where $\mathbf1\{\cdot\}$ denotes the indicator function. This explicit power model allows us to isolate the effects of sleep scheduling and switching overhead.
\subsection{Problem Formulation}\label{sec:problem}
The objective of the controller is to achieve a low average energy consumption of the cell while preserving the requested rate and playback availability of the clients. To measure the energy performance of a control policy $\pi$, we compare its energy consumption with that of always-on operation (i.e., $a_t=1$ for all $t$) under the same initial condition, and the average energy saving is defined as follows:
\begin{equation}
\overline{\mathrm{ES}}^{\pi}=\mathbb{E}\left[1-\frac{\sum_{t}E_t^{\pi}}{\sum_{t}E_t^{\rm AO}}\right],
\label{eq:saving}
\end{equation}
where $E_t^{\pi}$ and $E_t^{\rm AO}$ denote the energy consumption in~\eqref{eq:energy} under policy $\pi$ and under always-on operation, respectively, and the expectation is taken over the episodes. Then, the carrier sleep control problem can be formulated as follows:
\begin{maxi!}|s|
{\pi}{\overline{\mathrm{ES}}^{\pi}}{\label{eq:problem}}{\textbf{P1:}}
\addConstraint{\overline R^{\rm AO}-\overline R^{\pi}\le\epsilon_R}\label{eq:rate-constraint}
\addConstraint{\overline s^{\pi}/\Delta\le\epsilon_s}\label{eq:stall-constraint}
\addConstraint{a_t\in\{0,1\},\quad\forall t}\label{eq:binary-constraint}
\addConstraint{\text{\eqref{eq:available}--\eqref{eq:client}},}\label{eq:dynamics-constraint}
\end{maxi!}
where $\overline R^{\rm AO}$ and $\overline R^{\pi}$ denote the average requested rates under always-on operation and policy $\pi$, respectively, $\overline s^{\pi}$ is the average unavailable-playback duration under policy $\pi$, and the averages are taken over clients, time, and evaluation episodes. Constraint~\eqref{eq:rate-constraint} limits the requested-rate loss compared with always-on operation, while constraint~\eqref{eq:stall-constraint} limits the fraction of unavailable playback. Constraint~\eqref{eq:binary-constraint} indicates the binary carrier command, and finally, constraint~\eqref{eq:dynamics-constraint} represents the service, buffer, and client dynamics. Since these criteria are defined over averages, they are experiment-level acceptance criteria, instead of universal quality-of-experience (QoE) thresholds or per-user guarantees. It can be observed that P1 is a constrained sequential decision problem, in which the current carrier command affects the future requests of the clients through the feedback loop in~\eqref{eq:capacity}-\eqref{eq:client}. Solving P1 directly is challenging for two reasons. First, the adaptation function $f_{\rm client}(\cdot)$ is unknown to the controller. Secondly, the client states are only observed through delayed and lossy reports. Hence, the controller cannot directly evaluate the objective and constraints of P1 for a candidate decision. To address this problem, we propose EchoWM in the next section, which learns the unknown client response and plans the carrier commands based on the predicted application feedback.

\section{Proposed EchoWM Framework}\label{sec:proposed}
As shown in Fig.~\ref{fig:framework}, the proposed EchoWM consists of two main components: a structured application world model and a receding-horizon planner. The world model separates the known service mechanics from the unknown application response, which is learned from the observed transitions. Based on the world model, the planner evaluates candidate sleep schedules and selects the carrier command for each interval. We describe each component in the following subsections.

\subsection{Structured Application World Model}\label{sec:wm}
The proposed EchoWM combines the known updates with a learned approximation of the client adaptation function in~\eqref{eq:client}. To this end, we first construct the normalized input feature of client $u$ as follows:
\begin{equation}
z_{u,t}=\left[b_{u,t+1}/20,\ m_{u,t+1}/12,\ R_{u,t}/8,\ g_u/0.55\right].
\end{equation}
Then, a shared neural network $g_\theta$ with parameters $\theta$ predicts the rate increment of the client, and the next requested rate is estimated as follows:
\begin{equation}
\widehat R_{u,t+1}=\operatorname{clip}\!\left(R_{u,t}+g_\theta(z_{u,t}),R_{\min},R_{\max}\right),
\label{eq:neural}
\end{equation}
where $R_{\min}$ and $R_{\max}$ denote the minimum and maximum requested rates, respectively. By sharing the response model across clients while conditioning it on their profiles, we avoid training a separate network for each client. The carrier command affects $g_\theta$ indirectly, since the command changes the next buffer and service inputs through the known equations. Therefore, the complete transition of EchoWM remains action-conditioned.

To train the response model, we use only the client reports that are delivered to the controller. Specifically, a one-step training pair is formed when the reports of two consecutive source intervals both arrive, while the intervals with missing reports are excluded. For each pair, the next reported buffer and service estimate, the previous reported requested rate, and the client profile define the input feature $z$, and the difference between the two reported rates is used as the label. Then, we train $g_\theta$ by minimizing the squared error of the rate increment as follows:
\begin{equation}
\mathcal L(\theta)=\frac1{|\mathcal D|}\sum_{(z,\delta R)\in\mathcal D}
\left(g_\theta(z)-\delta R\right)^2,
\end{equation}
where $\mathcal D$ denotes the training dataset and $\delta R$ is the observed rate increment. The main advantage of separating the known mechanics from the unknown response is that it reduces the learning problem. In particular, the network does not need to rediscover the known mechanics, i.e., energy accounting, capacity limits, and buffer conservation, and it only learns the application decision that closes the feedback loop. For model selection, we use held-out collection episodes, so that model fitting and validation rely on transitions from disjoint episodes.

\subsection{Receding-Horizon Planning with Application Feedback}\label{sec:planning}
At each decision interval, the planner first reconstructs the current application state from the latest report by replaying the observed carrier-command history through the learned transition~\cite{karamzade2024reinforcement,pmlr-v331-karamzade26a}. Then, it evaluates a finite set $\mathcal A_H$ of binary command sequences over a horizon of $H$ intervals. For each sequence, the planner repeatedly computes the service, advances the buffers and service estimates, predicts the next requested rates, and uses those rates to generate the next requests. Accordingly, the score of a candidate sequence $\boldsymbol a$ is calculated as follows:
\begin{align}
J(\boldsymbol a)={}&\sum_{k=0}^{H-1}\left[
\frac{\widehat E_{t+k}}{E_{\rm ref}}+\ell_{t+k}\right]\notag\\
&+\eta\frac1U\sum_u[b_{\rm res}-\widehat b_{u,t+H}]_+,
\label{eq:planning}
\end{align}
where $\widehat E_{t+k}$ is the predicted interval energy obtained from~\eqref{eq:energy}, and the predicted service cost is given by:
\begin{equation}
\ell_j=\frac{\lambda}{U}\sum_u\left(1-\frac{\widehat R_{u,j+1}}{R_{\max}}\right)
+\frac{\mu}{U}\sum_u\frac{\widehat s_{u,j}}{\Delta}.
\end{equation}
Here, $E_{\rm ref}=\Delta(P_{\mathrm P}+P_{\mathrm A})$ is the always-on reference, $\lambda$ and $\mu$ are the weights of the rate and stall costs, respectively. The last term in~\eqref{eq:planning} is a terminal reserve with weight $\eta$ and reserve level $b_{\rm res}$. This terminal reserve discourages the planner from leaving nearly empty buffers just beyond the planning horizon. 
The rate weight $\lambda$ is selected on separate control-validation episodes, such that constraints~\eqref{eq:rate-constraint} and~\eqref{eq:stall-constraint} of P1 are satisfied. In this way, the planner handles the service constraints of P1 through weighted penalties, while the validation selection keeps the resulting operating point within the declared criteria.

\begin{algorithm}[t]
\caption{EchoWM Training \& Carrier Sleep Management}
\label{alg:echowm}
\begin{algorithmic}[1]
\State \textbf{Input:} Candidate set $\mathcal A_H$, horizon $H$, weights $\mu$ and $\eta$, and reserve level $b_{\rm res}$
\State Collect delivered client reports under exploratory carrier commands
\State Fit $g_\theta$ and select a checkpoint on held-out collection episodes
\State Select $\lambda$ on separate control-validation episodes
\For{each management interval}
\State Read the latest client report, report age, profiles, and action history
\State Reconstruct the current client state through the learned transition
\For{each candidate sequence $\boldsymbol a\in\mathcal A_H$}
\State Initialize the imagined states from the reconstructed state
\For{$k=0,\ldots,H-1$}
\State Update the capacity, service, buffers, and service estimates
\State Predict the client rates with $g_\theta$ and form the next requests
\EndFor
\State Calculate $J(\boldsymbol a)$ using~\eqref{eq:planning}
\EndFor
\State Execute the first command of the lowest-score sequence
\EndFor
\State \textbf{Output:} Carrier command $a_t$ at each management interval
\end{algorithmic}
\end{algorithm}

Following the receding-horizon principle, the controller only executes the first command of the sequence with the lowest score, and then repeats the planning at the next interval. Since the planning starts from the reconstructed state, the controller can replan at every interval even if the next client report is delayed or dropped. The candidate set $\mathcal A_H$ contains constant operation, all one-switch sequences, and periodic active/sleep patterns whose active and sleep periods last from one to four intervals. With this finite set, the planner can evaluate all candidates exhaustively at each interval. The overall procedure of EchoWM is summarized in Algorithm~\ref{alg:echowm}. The complexity of Algorithm~\ref{alg:echowm} mainly depends on state reconstruction and candidate evaluation. Let $K=|\mathcal A_H|$ and $P_\theta$ denote the number of candidate sequences and network parameters, respectively. Each predicted interval requires $\mathcal O(U\log U)$ for service allocation and $\mathcal O(UP_\theta)$ for client-response prediction. Since reconstruction requires at most $M$ intervals and candidate evaluation requires $KH$ intervals, the online complexity per management interval is $\mathcal O \left((M+KH)U(P_\theta+\log U)\right).$ Additionally, training for $N_{\rm ep}$ epochs on $|\mathcal D|$ client-transition pairs requires $\mathcal O(N_{\rm ep}|\mathcal D|P_\theta)$ operations.

\section{Performance Evaluation}\label{sec:evaluation}
In this section, we conduct a series of simulations to evaluate the performance of the proposed EchoWM and compare it with other benchmark schemes. First, we describe the simulation settings and the training process. Then, we introduce the benchmark schemes and the evaluation protocol, and finally, we present the numerical results.

\subsection{Simulation Settings}\label{sec:settings}
We consider a single cell with one BS and eight streaming clients and the detailed simulation parameters are provided in Table~\ref{tab:setup}. At the beginning of each episode, the initial buffers of the clients are uniformly sampled from [8, 18]~s, the requested rates start at 8~Mbit/s, and the service estimates are uniformly sampled from [7.5, 10.5]~Mbit/s. The capacity factor $\zeta$ follows a uniform distribution on [0.9, 1.1], and it remains fixed within each episode of 300~s. Since the clients start with buffered content, which represents established streams, our experiments focus on the ongoing service instead of the startup phase.

\begin{table}[t]\centering\footnotesize
\caption{Simulation Parameters}\label{tab:setup}
\def\celltext#1{{\raggedright #1\par}}
\begin{tabular}{@{}p{.39\columnwidth}p{.56\columnwidth}@{}}\toprule
Parameter & Value\\\midrule
\celltext{Clients} & \celltext{8; $g_u\in\{0.15,0.35,0.55\}$}\\
\celltext{Decision interval / episode} & \celltext{1 s / 300 s}\\
\celltext{PCell / SCell capacity} & \celltext{24 / 72 Mbit/s, scaled by $\zeta$}\\
\celltext{Capacity factor $\zeta$} & \celltext{Uniform on [0.9, 1.1], fixed per episode}\\
\celltext{PCell / active / sleep power} & \celltext{40 / 100 / 10 W}\\
\celltext{$\alpha$  / $M$ / $\beta$ / $\gamma$ / $\mu$ / $\eta$ / $b_{\rm res}$} & \celltext{0.25 / 16 / 0.3 / 0.7 / 20 / 0.2 / 8s}\\

\celltext{Wake fraction / switch energy} & \celltext{0.2 / 5 J}\\
\celltext{Buffer target / refill time} & \celltext{20 / 4 s}\\
\celltext{Requested-rate range} & \celltext{1-8 Mbit/s, continuous}\\
\celltext{Reports} & \celltext{4-s delay; 25\% drops; max age 16 s}\\
\celltext{Prediction horizon} & \celltext{12 s; 1 s in the ablation}\\
\celltext{Model} & \celltext{Shared 4-48-48-1 tanh network}\\
\celltext{Real interaction budgets} & \celltext{3,000 / 9,000 / 24,000 cell slots}\\
\celltext{Acceptance criteria} & \celltext{Rate loss $\le0.10$ Mbit/s; stall fraction $\le0.001$}\\\bottomrule
\end{tabular}
\vspace{-0.1in}
\end{table}

To emulate the rate adaptation of clients, the simulator adopts a transparent smooth-hysteresis ABR rule. Let $q=\operatorname{clip}(0.95m,R_{\min},R_{\max})$ denote the reference rate derived from the recent service and $\sigma(y)=1/(1+e^{-y})$ denote the sigmoid function. Then, the response of the client is given by:
\begin{equation}
f_{\rm client}(b,m,R;g_u)=\operatorname{clip}\!\left(R+g_uh(b,q-R)(q-R),1,8\right),
\end{equation}
where $h(b,\delta)=\sigma((8-b)/1.2)$ for $\delta<0$ and $h(b,\delta)=\sigma((b-12)/1.2)$ otherwise. With this rule, a low buffer allows the client to adapt downward, whereas a larger buffer supports the recovery of the requested rate. Additionally, each client independently receives one of the three visible profiles, i.e., $g_u=0.15$, 0.35, and 0.55, with probabilities of 0.34, 0.32, and 0.34, respectively. One noteworthy point is that we adopt this rule to instantiate the feedback mechanism in our simulator, rather than to represent a production ABR implementation. Furthermore, only the simulator calls this rule, while the planner of our proposed EchoWM relies on its learned approximation in~\eqref{eq:neural}.

To train the response model, we collect 80 exploratory episodes of 300 intervals, in which randomly chosen carrier commands are held for 1 to 15~s. The first 3,000, 9,000, and 24,000 real cell slots of this collection provide the training data for the three interaction budgets. As described in Section~\ref{sec:wm}, only consecutive delivered report pairs enter the regression dataset, and at the full budget, 13,271 such pairs yield 106,168 client rows. For each budget, we split the complete collection episodes into 80\% for model fitting and 20\% for checkpoint selection. Then, three network initializations share each collection prefix, and each of them is trained for 60 epochs using the Adam optimizer~\cite{kingma2015adam} with a learning rate of 0.002 and a batch size of 2,048. For the RL benchmarks, we conduct three separate online training runs for each budget. All neural models are trained on an NVIDIA RTX 3060 Ti GPU. 

\subsection{Benchmarks for Performance Comparison}\label{sec:benchmarks}
To evaluate the effectiveness of the proposed EchoWM, we compare its performance against the following benchmarks:
\begin{itemize}
\item \textit{Always-on}: The SCell remains active in every interval. This scheme serves as the reference for energy saving and rate loss, and it is also used as the fallback when no setting satisfies the acceptance criteria.
\item \textit{Utilization rule}: This rule activates the SCell when the offered traffic exceeds a multiple of the PCell capacity, where the multiple is selected from 2 to 4.
\item \textit{Buffer rule}: This rule activates the SCell when the minimum client buffer falls below a threshold, and puts it to sleep after the minimum client buffer exceeds that threshold by two seconds.
\item \textit{DQN~\cite{mnih2015human} and PPO~\cite{schulman2017proximal}}: These model-free RL policies use two hidden layers with 64 units each, and they are trained on the common delayed-report observation with 52 elements and the same carrier-cost reward. During training, each policy immediately receives the scalar reward at every interval, while the policy only accesses this observation at decision time.
\item \textit{Recurrent PPO}: This model-free policy adopts a long short-term memory (LSTM) network~\cite{hochreiter1997long} with the same reported values, action history, and training reward as DQN and PPO. Moreover, its recurrent state can summarize the observations from earlier intervals.

\end{itemize}

\begin{figure*}[t]\centering
\includegraphics[width=0.9\textwidth]{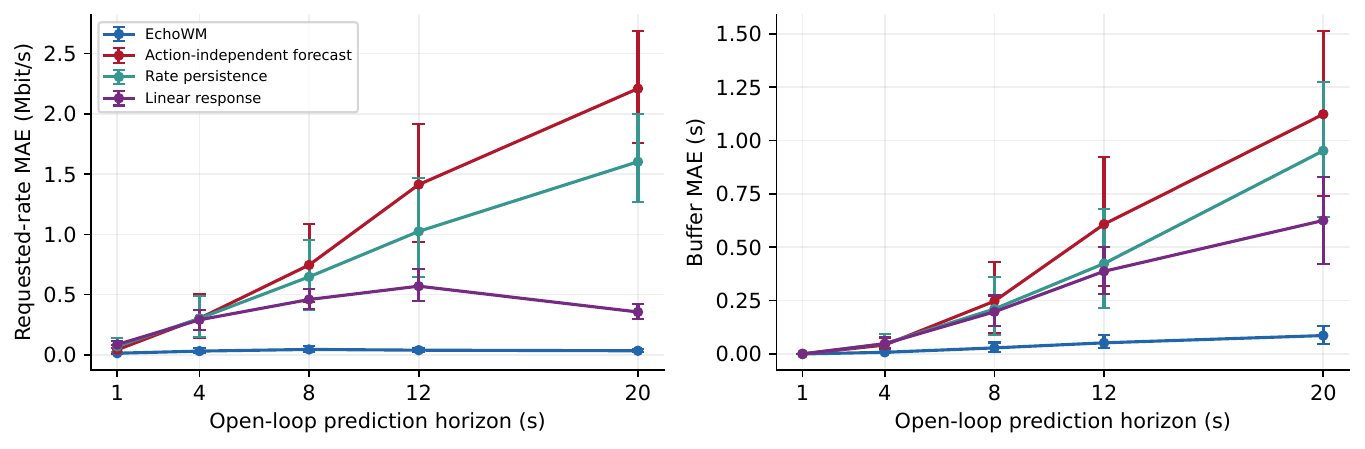}
\caption{Open-loop prediction errors of the requested rate (left) and buffer (right) for four predictors on held-out client trajectories under four prespecified carrier schedules in the homogeneous diagnostic setting.}
\vspace{-0.1in}
\label{fig:prediction}\end{figure*}
\subsection{Evaluation Protocol}\label{sec:selection}
At each interaction budget, we tune all controllers on 12 validation episodes, selecting the highest energy saving that satisfies constraints~\eqref{eq:rate-constraint} and~\eqref{eq:stall-constraint}. We then evaluate on 40 paired test episodes, reporting energy saving, requested-rate loss, and unavailable playback. Each cell interval counts as one interaction, and ablations retain EchoWM's selected rate weight at the full budget.

\subsection{Results Analysis}\label{sec:results}

\begin{table}[t]\centering\footnotesize
\caption{Client Response to Initial SCell Sleep Pulses}\label{tab:pulse}
\begin{tabular}{@{}rrrrrr@{}}\toprule
& \multicolumn{3}{c}{Rate loss (Mbit/s)} & \multicolumn{2}{c}{Unavailable playback (\%)}\\
Sleep (s) & Slow & Std. & Fast & Slow & Fixed\\\midrule
0  & 0.000 & 0.000 & 0.000 & 0.000 & 0.000\\
4  & 0.087 & 0.097 & 0.103 & 0.000 & 0.000\\
8  & 0.311 & 0.300 & 0.305 & 0.000 & 0.000\\
12 & 0.645 & 0.587 & 0.594 & 0.000 & 0.000\\
16 & 1.043 & 0.934 & 0.940 & 0.000 & 0.370\\
24 & 1.819 & 1.647 & 1.632 & 0.002 & 4.070\\
40 & 3.092 & 2.993 & 2.975 & 0.003 & 19.959\\
\bottomrule\end{tabular}\par\smallskip\scriptsize
\vspace{-0.10in}
\end{table}

\subsubsection{Impact of Carrier Sleep on Adaptive Clients}
We first verify the application feedback that our proposed EchoWM aims to capture. To this end, we use two diagnostics from a homogeneous implementation of the carrier-sleep task, whose seeds are disjoint from those of the controller tests. In Table~\ref{tab:pulse}, we put the SCell to sleep for a fixed initial duration and then restore the service over a common 60-s observation window. The requested-rate loss of the adaptive clients increases with the sleep duration, while their playback remains almost uninterrupted. Specifically, a 24-s sleep pulse lowers the mean requested rate by 1.647~Mbit/s for the standard adaptive clients (i.e., $g_u=0.35$), 1.819~Mbit/s for the clients with slower adaptation (i.e., $g_u=0.15$), and 1.632~Mbit/s for the clients with faster adaptation (i.e., $g_u=0.55$). On the other hand, the clients with a fixed rate of 8~Mbit/s keep their requested rate, but they experience 4.07\% unavailable playback under the same pulse. These results confirm that a sleep decision changes the delivered service and, consequently, the future requests of adaptive clients, so that the service degradation appears as a lower requested rate instead of playback interruptions.

\begin{table}[t]\centering\footnotesize
\caption{Energy Saving and Requested-Rate Loss of All Controllers on the Paired Test at Three Real-Interaction Budgets}\label{tab:learning}
\begin{tabular}{@{}lrrrrrr@{}}\toprule
& \multicolumn{2}{c}{3,000 slots} & \multicolumn{2}{c}{9,000 slots} & \multicolumn{2}{c}{24,000 slots}\\
Controller & ES (\%) & Loss & ES (\%) & Loss & ES (\%) & Loss\\\midrule
Always-on & 0.00 & 0.000 & 0.00 & 0.000 & 0.00 & 0.000\\
Utilization rule & 22.44 & 0.003 & 22.44 & 0.003 & 22.44 & 0.003\\
Buffer rule & 26.19 & 0.086 & 26.19 & 0.086 & 26.19 & 0.086\\
EchoWM & 23.83 & 0.058 & 24.07 & 0.062 & 24.02 & 0.059\\
DQN & 15.83 & 0.052 & 23.36 & 0.058 & 24.43 & 0.070\\
PPO & 0.00 & 0.000 & 14.98 & 0.005 & 19.83 & 0.096\\
Recurrent PPO & 0.13 & 0.001 & 3.48 & 0.000 & 6.07 & 0.000\\
\bottomrule\end{tabular}
\vspace{-0.1in}
\end{table}

\subsubsection{Accuracy of the Learned Application Response}
In Fig.~\ref{fig:prediction}, we evaluate the open-loop prediction accuracy of four predictors on held-out client trajectories from the same homogeneous diagnostic. Each predictor forecasts the future requested rates and buffers under four prespecified carrier schedules. Besides EchoWM and the action-independent forecast, the rate-persistence predictor keeps the current requested rates unchanged, while the linear-response predictor replaces $g_\theta$ with a fitted linear model of the same input features. Overall, EchoWM outperforms the three benchmark predictors at every horizon. In contrast, the requested-rate mean absolute error (MAE) of EchoWM stays at or below 0.045~Mbit/s for all horizons. Specifically, at the 12-s horizon, the MAE of EchoWM is 0.039~Mbit/s, whereas the action-independent forecast, rate persistence, and linear response obtain 1.413, 1.025, and 0.570~Mbit/s, respectively. These results show that the learned application transition captures how carrier commands propagate to the future requests. However, an accurate prediction alone is insufficient to guarantee better carrier control, which is evaluated in the following.

\subsubsection{Sample Efficiency under Different Interaction Budgets}
In Table~\ref{tab:learning}, we compare all controllers on the frozen test of the heterogeneous, delayed-report scenario, where each budget counts the real cell slots used for training. As the budget grows, the held-out report-pair root mean square error (RMSE) of the learned response decreases from 0.127~Mbit/s (3,000 slots) to 0.071~Mbit/s (9,000 slots) and 0.033~Mbit/s (24,000 slots). Nevertheless, our proposed EchoWM already saves 23.83\% of the energy with a mean requested-rate loss of 0.058~Mbit/s at the smallest budget, whereas DQN saves 15.83\%. PPO and recurrent PPO obtain even lower accepted savings at this budget. The mean paired saving difference between EchoWM and DQN is 8.00 percentage.

\begin{table}[t]\centering\footnotesize
\caption{Performance of EchoWM Ablations at the 24,000 Slots}\label{tab:ablation}
\begin{tabular}{@{}lrrc@{}}\toprule
Planner & ES (\%) & Rate loss & Meets limit\\\midrule
EchoWM & 24.02 & 0.059 & Yes\\
No reconstruction & 27.58 & 0.186 & No\\
Action independent & 64.24 & 5.042 & No\\
One-step & 64.27 & 5.045 & No\\
\bottomrule\end{tabular}
\vspace{-0.1in}
\end{table}

When the budget increases to 9,000 slots, EchoWM saves 24.07\%, compared with 23.36\% for DQN. With 24,000 slots, DQN saves 24.43\% and slightly surpasses EchoWM (24.02\%), and the paired EchoWM-minus-DQN difference is $-$0.41 percentage. Thus, the measured benefit of EchoWM is concentrated in the low-data regime, and DQN catches up as more real interactions become available. This trend can be explained by the structure of EchoWM, which only learns the client response and takes the remaining dynamics from the known equations, whereas the model-free policies have to learn the whole closed-loop behavior from the interactions. Additionally, the energy saving of EchoWM remains nearly unchanged across the three budgets, which suggests that the response model trained with 3,000 slots is already accurate enough for planning in this scenario. Hence, this low-dimensional rule is a strong operational benchmark, and we further examine its behavior under different report delays in the following.

\subsubsection{Effect of State Reconstruction and Action-Dependent Prediction}
To understand which components allow the proposed EchoWM to preserve the requested rate, we change one component of EchoWM at the full budget of 24,000 slots while retaining its selected rate weight, as shown in Table~\ref{tab:ablation}. When the planner starts from the stale report without replaying the carrier commands, the raw energy saving increases to 27.58\%, but the requested-rate loss also increases to 0.186~Mbit/s, which exceeds the declared limit of 0.10~Mbit/s. This result can be attributed to the fact that the stale report misses the effects of the carrier commands issued after it, and thus, the planner tends to overestimate the remaining service margin of the clients. As evidence, the SCell sleeps in 43.4\% of the intervals without reconstruction, compared with 39.1\% for EchoWM. Similarly, the action-independent and one-step planners achieve high raw savings of 64.24\% and 64.27\% by allowing the requests to collapse, and they fail the service criterion with rate losses of 5.042 and 5.045~Mbit/s, respectively. Specifically, the action-independent forecast assigns the same rate penalty to every candidate schedule. On the other hand, the one-step planner only sees the consequence of a sleep decision within a single interval, although the clients mainly reduce their requested rates after their buffers have been depleted over several intervals. As a result, neither of them can recognize the rate degradation caused by carrier sleep, and they keep the SCell asleep in 99.9\% and 100.0\% of the intervals, respectively.

\subsubsection{Robustness against Report Delays}
In Table~\ref{tab:delay}, we evaluate the full-budget policies under different report delays on a separate split of 30 paired episodes, without retraining or retuning. In this experiment, only the report delay changes, while the report-drop probability remains at 25\%, and all policies keep the settings selected at the 4-s delay (i.e., the delay of the main test). As we can see, the energy saving and requested-rate loss of EchoWM remain nearly constant when the delay increases from 0 to 8~s, which can be attributed to the state reconstruction that compensates for the age of the latest report. In contrast, the raw energy saving of the tuned buffer rule increases with the delay; however, its mean rate loss reaches 0.350~Mbit/s at 8~s and violates the service limit, while the rate loss of EchoWM is only 0.054~Mbit/s at the same delay. This is mainly because the buffer rule reacts to the reported minimum buffer, which becomes more outdated as the delay grows. DQN remains competitive in energy saving during this sweep. Its rate loss rises to 0.093~Mbit/s at 8~s. These results support our state reconstruction as a way to preserve the service under stale feedback.

\begin{table}[t]\centering\footnotesize
\caption{Zero-Shot Performance under Different Report Delays}\label{tab:delay}
\begin{tabular}{@{}rrrr@{}}\toprule
Delay (s) & EchoWM & DQN & Buffer\\\midrule
0 & 24.90/0.053 & 24.95/0.074 & 25.61/0.008\\
4 & 24.91/0.054 & 25.41/0.069 & 27.11/0.084\\
8 & 24.91/0.054 & 25.37/0.093 & 28.51/0.350\\
\bottomrule\end{tabular}
\vspace{-0.1in}
\end{table}

\vspace{-0.1in}
\section{Conclusion}\label{sec:conclusion}
In this paper, we investigate the carrier sleep control problem, in which adaptive video clients respond to capacity reductions by lowering their requested rates, and therefore, the service degradation is hidden from the traffic observed at the base station. To address this problem, we propose EchoWM, a structured world model that learns the client response from delivered report pairs and combines it with the known capacity, buffer, and energy mechanics of the cell. When client reports are delayed or missing, EchoWM reconstructs the present client state through its own action history, and then a receding-horizon planner selects the carrier commands based on the predicted application feedback. Simulation results show that EchoWM achieves a higher mean energy saving than the tested DQN, PPO, and recurrent PPO policies under the same service limits with limited real interactions. Additionally, the ablation studies demonstrate the importance of state reconstruction and action-dependent prediction in preserving the requested rate. EchoWM maintains stable performance under longer report delays, while the buffer rule violates the rate-loss constraint. These findings highlight the benefits of structured application modeling for energy-efficient network control under limited interactions and delayed feedback.
\bibliographystyle{IEEEtran}
\bibliography{EchoWM_references}
\end{document}